# Natural pigments from microalgae grown in industrial wastewater

Larissa T. Arashiro[a,b], María Boto-Ordóñez[c], Stijn W.H. Van Hulle[b], Ivet Ferrer[a], Marianna Garfí[a]*, Diederik P.L. Rousseau[b]

[a]GEMMA - Group of Environmental Engineering and Microbiology, Department of Civil and Environmental Engineering, Universitat Politècnica de Catalunya · BarcelonaTech, c/ Jordi Girona 1-3, Building D1, 08034 Barcelona, Spain

[b]LIWET – Laboratory of Industrial Water and Ecotechnology, Department of Green Chemistry and Technology, Ghent University Campus Kortrijk, Graaf Karel de Goedelaan 5, 8500 Kortrijk, Belgium

[c]KASK, School of Arts, University College Ghent, 9000 Ghent, Belgium

*Corresponding author: Tel: +34 9340 16412
*E-mail address*: marianna.garfi@upc.edu (M. Garfí).

**Abstract**

The aim of this study was to investigate the cultivation of *Nostoc* sp., *Arthrospira platensis* and *Porphyridium purpureum* in industrial wastewater to produce phycobiliproteins. Initially, light intensity and growth medium composition were optimized, indicating that light conditions influenced the phycobiliproteins production more than the medium composition. Conditions were then selected, according to biomass growth, nutrients removal and phycobiliproteins production, to cultivate these microalgae in food-industry wastewater. The three species could efficiently remove up to 98%, 94% and 100% of COD, inorganic nitrogen and $PO_4^{3-}$-P, respectively. Phycocyanin, allophycocyanin and phycoerythrin were successfully extracted from the biomass reaching concentrations up to 103, 57 and 30 mg/g dry weight, respectively. Results highlight the potential use of microalgae for industrial wastewater treatment and related high-value phycobiliproteins recovery.

## 1. Introduction

Microalgae are known to have a great capacity to efficiently utilize nutrients from wastewaters, since their cultivation requires high amounts of nitrogen and phosphorus. Besides enabling efficient wastewater treatment, microalgae biomass is a potential source of valuable chemicals and other products, attracting wide interest lately (Vuppaladadiyam et al., 2018). The phycobiliproteins (PBPs) are among those chemicals and are exclusive to the cyanobacteria, red algae, and the Cryptophyta and Glaucophyta (Borowitzka, 2013). These molecules are auxiliary pigments, water soluble and highly fluorescent proteins with linear prosthetic groups (bilins) that are linked to specific cysteine residues. Depending on their composition and content of chromophores, PBPs may be classified as phycocyanins (PC, $\lambda_{max}$ = 610-625 nm), phycoerythrins (PE, $\lambda_{max}$ = 490-570 nm), and allophycocyanins (APC, $\lambda_{max}$ = 650-660 nm) (Noreña-caro and Benton, 2018).

The main commercial producers of phycobiliproteins are the cyanobacterium *Arthrospira* and the rhodophyte *Porphyridium* (Spolaore et al., 2006). Furthermore, the cyanobacterium *Nostoc* has also been recently put forward as a potential source of phycobiliproteins (Johnson et al., 2014). Besides playing an important role in the pigmentation metabolism of microalgae, phycobiliproteins also exhibit some useful biological functions, such as antioxidative, anticarcinogenic, anti-inflammatory, antiangiogenic, and neuro and hepatoprotective (Cuellar-Bermudez et al., 2015). Phycobiliproteins have high commercial value as natural colorants in the nutraceutical, cosmetic, and pharmaceutical industries, as well as applications in clinical research and molecular biology (Chiong et al., 2016) and as natural dyes in the textile industries (Okolie et al., 2019). Recent studies have shown that extracts of red pigment from the macroalgae *Gracilaria vermiculophylla* and the blue pigment from the *Arthrospira platensis* showed even distribution on the cotton and wool fabrics, with results representing the viability and the quality of naturally dyed textiles (Ferrándiz et al., 2016; Moldovan et al., 2017).

Microalgae cultivation for chemical production using standard culture media can account for high costs. In order to improve the economic feasibility, these costs could be reduced by changing culture media concentrations, for instance through dilution (Delrue et al., 2017), or by using alternative components, including wastewater (Arashiro et al., 2019; Kumar et al., 2020; Van Den Hende et al., 2016b). Several studies have proposed alternatives to reduce costs for *Arthrospira* production, such as the use of fertilizers or seawater with and without enrichment of $NaHCO_3$ and $NaNO_3$ (Gami et al., 2011), swine wastewater (Yilmaz and Sezgin, 2014) and brine wastewater (Duangsri and Satirapipathkul, 2011; Volkmann et al., 2008). Moreover, utilizing seawater rather than freshwater for production of saline microalgae, such as *A. platensis*, would be practical and cost effective (Mahrouqi et al., 2015). The potential of the cyanobacterium *Nostoc* to promote wastewater treatment and biomass production has also been reported (El-Sheekh et al., 2014; Talukder et al., 2015).

The production of phycobiliproteins from biomass grown in wastewater has not been extensively explored. In this sense, this study investigated the performance of unialgal cultivations and phycobiliproteins production of *Nostoc* sp., *A. platensis* and *Porphyridium purpureum* grown in: i) standard growth media with varied compositions and different light intensities (in view of optimization of the reactor conditions); and ii) food-industry wastewater under optimal light conditions. As such, this is the first study describing the possibility to recover natural pigments from these three species using effluent in order to reduce the high costs of its production (since nutrients are recovered from wastewater and therefore no chemicals would be needed for growth medium). Not only the feasibility is assessed, but also different aspects for pigments production by these microalgae under varied cultivation conditions.

**2. Materials and methods**

*2.1. Inocula and culture conditions*

A microalgae sample of *Nostoc* sp. was obtained from the Spanish Bank of Algae, University of Las Palmas de Gran Canaria (Spain) and samples of *A. platensis* and *P. purpureum* were obtained from the laboratory of the School of Arts (KASK), University College Ghent (Belgium). The microalgae species were cultured in Erlenmeyer flasks containing sterilized growth media in a room with constant temperature of 22°C and continuously mixed with magnetic stirrers (Hei-mix S, Heidolph, Germany) at 200 rpm. *Nostoc* sp. was cultivated in BG-11 medium (Stanier et al., 1971). *A. platensis* was cultivated in a modified Zarrouk's medium (Castro et al., 2015). *P. purpureum* was cultivated in a modified artificial sea water (ASW) (Velea et al., 2011).

*2.2. Optimization experiment*

The optimization experiment was carried out to find optimal conditions of light intensity and medium composition to reach high phycobiliproteins yields. The cultivations were done in 750 mL Erlenmeyer flasks filled with 600 mL culture media. The flasks were maintained in a room with constant temperature of 22°C and aeration tubes providing approximately 3 L air/min were used to provide constant mixing and inorganic carbon. Illumination was provided by cool-white fluorescent lamps (Lumilux L 36W/840,

Osram, Germany) with a light:dark cycle of 14:10 h. Two growth parameters were tested for each species: light intensity and concentration of the main ingredient of standard growth medium: $NaNO_3$ for *Nostoc* sp., $NaHCO_3$ for *A. platensis* and NaCl (as sea salt) for *P. purpureum*. Varying cultivation conditions tested as a $3^2$ full factorial design (for each species) are described in Table 1.

*2.3. Wastewater treatment experiment*

The set-up consisted of five photobioreactors, one for *Nostoc* sp., two for *A. platensis* and two for *P. purpureum*, made with 90 µm thick bags composed of polyamide and polyethylene (Sacs sous vide, Spain) with working volume of 8L each. The wastewater used in this experiment was an upflow anaerobic sludge blanket (UASB) effluent from a food company that markets plant-based products (Alpro, Wevelgem, Belgium). Since the wastewater has reduced organic matter concentrations (after UASB), while still containing sufficient nutrients, this effluent is promising for the production of microalgal biomass, as previously described by Van Den Hende et al. (2016).

The wastewater was filtered using 0.45 µm membrane filters to remove any suspended particles before being used in the experiment. The filtered wastewater was refrigerated at 4°C to prevent any biochemical process that could change its composition. The photobioreactors were fed with the wastewater mixed with: i) standard medium BG-11 for *Nostoc* sp., at ratio 50% WW + 50% BG-11; and ii) artificial sea water (ASW) for saline species *A. platensis* and *P. purpureum*, at two different volume ratios: 50% WW + 50% ASW and 75% WW + 25% ASW. The ratios of wastewater tested were defined based on the results of biomass growth and phycobiliproteins production of the optimization experiment.

*2.4. Biomass growth rate determination*

During the optimization experiment, biomass growth was monitored by withdrawing aliquots of 3 mL of culture media every 2 to 4 days and measuring the optical density (OD) at 680nm. Considering that working volume in this experiment was low (600 mL), biomass growth was monitored through OD to avoid withdrawing large volumes from vessels. This was done since good correlations between OD and dry weight (DW) were observed: $R_2 = 0.989$ (*Nostoc* sp.), $R_2 = 0.999$ (*A. platensis*) and $R_2 = 0.993$ (*P. purpureum*). On the other hand, during the wastewater experiment, as the working volume was 8 L, biomass growth was monitored through volatile suspended solids (VSS) measurement by filtering 50 mL of culture media.

Maximum specific growth rate ($\mu_{max}$) was calculated through Eq. 1, considering biomass concentrations during the exponential phase (Andersen, 2005).

$$\mu_{max} = \frac{\ln\left(\frac{C_f}{C_i}\right)}{(t_f - t_i)} \qquad \text{Eq. 1}$$

where $C_i$ and $C_f$ are the biomass concentrations at time initial ($t_i$) and final ($t_f$), respectively.

*2.5. Analytical methods*

Total suspended solids (TSS), volatile suspended solids (VSS) were determined according to Standard Methods (APHA-AWWA-WEF, 2012) and chemical oxygen demand (COD) was determined according to the method for high salinity proposed by Kayaalp et al. (2010). The measurement of total nitrogen (TN), total phosphorus (TP), $NH_4^+$-N, $NO_3^-$-N, $NO_2^-$-N, $PO_4^{3-}$-P were done with spectrophotometric test kits (Hach,

USA). Total inorganic nitrogen (TIN) is expressed as the sum of $NH_4^+$-N, $NO_3^-$-N and $NO_2^-$-N. pH and electrical conductivity (EC) were measured with a portable multi-parameter meter HQ30d (Hach, USA). All the analyses were done in triplicate and results are given as average values and standard deviation.

*2.6. Phycobiliproteins extraction*

Phycobiliproteins content in the biomass grown during the optimization and the wastewater experiments was quantified. The culture medium was centrifuged at 1,160 g for 15 min (Hermle Z 300 K, Germany) and the biomass pellets were frozen (-21°C) until further use. Approximately 1 g of biomass was added into 15mL centrifuge tubes and sodium phosphate buffer (0.1M, pH 7) was used as solvent at a proportion 1:10 (w:w, biomass:solvent). The tubes were then submitted to two freeze-thawing (-21°C to 4°C in darkness) cycles. The resulting slurry was centrifuged at 9,500 g for 15 min at 4°C (Hermle Z 300 K, Germany) to remove the cell debris. The supernatant was collected and measured in a spectrophotometer (Shimadzu UV-1280, Japan) at 280, 562, 615 and 652 nm. The amounts of phycocyanin (PC), allophycocyanin (APC) and phycoerythrin (PE) were calculated according to Eq. 2, Eq. 3 and Eq. 4, respectively (Bennett and Bogobad, 1973).

$$PC\ (mg/mL) = [A_{615} - (0.474 * A_{652})] / 5.34 \qquad \text{Eq. 2}$$

$$APC\ (mg/mL) = [A_{652} - (0.208 * A_{615})] / 5.09 \qquad \text{Eq. 3}$$

$$PE\ (mg/mL) = [A_{562} - (2.41*PC) - (0.849*APC)] / 9.62 \qquad \text{Eq. 4}$$

where $A_{562}$, $A_{615}$ and $A_{652}$ are the absorbances measured at the respective wavelengths. Purity was determined as the absorbance ratios of $A_{620}/A_{280}$ for phycocyanin, $A_{652}/A_{280}$ for allophycocyanin and $A_{565}/A_{280}$ for phycoerythrin (Cuellar-Bermudez et al., 2015).

*2.7. Statistical analyses*

The average values measured during the experiments were analyzed using multi-factor analysis of variance (ANOVA) and the significant differences amongst treatments and phycobiliproteins production were determined using Fisher test at 95% confidence interval level. All statistical analyses were done using Minitab 18 (Minitab Inc., PA, USA).

**3. Results and discussion**

*3.1. Optimization experiment*

This experiment was carried out in order to define the best conditions of light intensity and growth media composition that would suggest the optimal combinations to maximize biomass growth and phycobiliproteins production of *Nostoc* sp., *A. platensis* and *P. purpureum*.

*3.1.1. Biomass growth and nutrients removal*

Biomass growth curves of *Nostoc* sp., *A. platensis* and *P. purpureum* during this experiment are shown in Figure 1. Growth curves and their respective statistical significance suggest that light conditions had more influence on biomass growth than the media composition (Figure 1). Initial and final concentrations of nutrients as TIN and $PO_4^{3-}$-P are shown in Figure 2. The three microalgae species could grow well in all trials until the end of the batch cultivation, except the L3C1 of both *Nostoc* sp. and *P. purpureum*, which indicates that high light conditions promoted fast growth in the initial phase of cultivation, and led to nutrients starvation faster than in other trials. This can be supported by the final TIN concentrations observed in these two trials shown in Figure 2.

3.1.1.1. *Nostoc* sp.

The lowest biomass concentration of *Nostoc* sp. after 20 days of cultivation was measured in L1C2 with 1.1 g/L, while the highest in L3C3 with 2.2 g/L. Increasing light conditions led to higher biomass growth (except for L3C1, as previously mentioned). This is in accordance with the results shown by Spencer et al. (2011), in which growth of *Nostoc spongiaeforme* increased up to 227 µE/m$^2$s (similar to L3 in this study). Initial and final TIN concentrations in the growth media of *Nostoc* sp. showed removal rates ranging from 3.8 to 7.6 mg N/Ld (28 to 94% removal efficiency). Although higher light intensities led to higher biomass growth, no significant difference was observed on TIN removal by *Nostoc* sp. when comparing different light intensities in L1, L2 and L3 (p-value = 0.238), as well as different initial NaNO$_3$ concentrations in C1, C2 and C3 (p-value = 0.422). Likewise, PO$_4^{3-}$-P removal rates ranged from 0.35 to 0.39 mg P/Ld (93 to 99% removal efficiency). Likewise, no significant difference was observed on phosphorus removal by *Nostoc* sp. when comparing different initial NaNO$_3$ concentrations in C1, C2 and C3 (p-value = 0.063) and different light intensities in L1, L2 and L3 (p-value = 0.568).

3.1.1.2. *A. platensis*

The cultivations of *A. platensis* reached the lowest biomass concentration after 20 days in L1C1 with 1.0 g/L, while the highest in L3C3 with 3.0 g/L. Likewise, Markou et al. (2012) reported that increasing light intensity from 24 to 60 µE/m$^2$s led to 2.7-fold higher *A. platensis* biomass production in standard growth medium. Castro et al. (2015) also reported that higher light conditions and $NaHCO_3$ concentrations increased *A. platensis* biomass production. Initial and final TIN concentrations in the growth media of *A. platensis* showed removal rates ranging from 6.7 to 13.3 mg N/Ld (52 to 100% removal efficiency). Removal rates of both medium and high light intensities were significantly higher than low light intensity (p-value = $4 \times 10^{-5}$), while different concentrations of $NaHCO_3$ in the growth media did not have affect TIN removal (p-value = 0.996). Regarding phosphorus removal, *A. platensis* showed removal rates ranging from 3.2 to 3.6 mg P/Ld (73 to 82% removal efficiency). When comparing the influence of light conditions, L1, L2 and L3 showed similar performances of phosphorus removal (p-value = 0.983), while when varying concentrations of $NaHCO_3$, C3 showed significantly higher removal rates than C1 and C2 (p-value = 0.004).

3.1.1.3. *P. purpureum*

The cultivations of *P. purpureum* showed the lowest biomass after 18 days of cultivation measured in L1C1 with 0.9 g/L, while the highest in L2C3 with 2.3 g/L. The results showed faster growth under high light intensities (L3), but over time, the cultivations under medium light intensities (L2) overtook, reaching the maximum biomass concentrations at the end of the experiment. Similar results were reported by Sosa-Hernández et al. (2019) with 0.3 L cultivations of *P. purpureum* in Bold 1NV and Erdshreiber media, where light intensity of 100 $\mu E\ m^{-2}\ s^{-1}$ (compared to 65 and 30 $\mu E\ m^{-2}\ s^{-1}$) resulted in higher biomass. Regarding NaCl concentrations, medium salinity level (C2) seemed to be more favorable for biomass growth, which was also reported by previous studies with the same species (Aizdaicher et al., 2014; Kathiresan et al., 2007).

Similarly to *A. platensis*, removal rates by *P. purpureum* ranged from 4.1 to 8.6 mg N/Ld (45 to 98% removal efficiency), in which L2 and L3 had significantly higher TIN removal than L1 (p-value = 0.039), while concentration of NaCl in the growth media did not have a significant influence on the TIN removal (p-value = 0.399). Regarding phosphorus removal, removal rates ranged from 0.56 to 0.79 mg P/Ld (80 to 100% removal efficiency), in which varying light conditions did not affect removal rates (p-value = 0.677), while higher concentrations of NaCl (both C2 and C3) removed better than low concentration (C1) (p-value = 0.011).

It is noteworthy, thus, to highlight that the higher light intensity did not improve the $PO_4^{3-}$-P removal in all trials, as observed with TIN removal (especially for *A. platensis* and *P. purpureum*). Considering the biomass stoichiometry, as $PO_4^{3-}$-P was not proportionally removed compared to TIN, the higher TIN removal was not related to biomass growth only, but possibly to the occurrence of nitrification-denitrification process during the cultivation period. This is in accordance with the increase in concentrations of $NO_2^-$-N in all trials during the experiment.

### *3.1.2. Phycobiliproteins*

The concentrations and yield of phycobiliproteins extracted from the biomass grown during the optimization experiment are shown in Figure 3. The concentrations represent the amount of phycobiliproteins per dry weight, while the yield represents the amount of phycobiliprotein per unit of volume and time in each trial, i.e. considering the content and biomass production during this period.

3.1.2.1. *Nostoc* sp.

Results of *Nostoc* sp. cultivations show that, in average, phycocyanin was the most abundant phycobiliprotein, followed by allophycocyanin and phycoerythrin. The maximum total concentration of phycobiliproteins was obtained in L1C3 (199 mg/g DW), while the minimum in L3C2 (15 mg/g DW). The concentrations observed in this study in accordance with previous results of *Nostoc* sp. cultivated in standard growth media (Khattar et al., 2015; Ma et al., 2015). Regardless of initial $NaNO_3$ concentrations, lower light intensities produced more phycobiliproteins, with higher purity observed under medium light intensity (L2) (Figure 3), which is in accordance with previous studies (Johnson et al., 2014; Ma et al., 2015). Regarding $NaNO_3$, varying its concentration did not influence phycobiliproteins content as much as varying light conditions, which is in accordance with the study carried out by Rosales Loaiza et al. (2016). Under low light intensity (L1), C3 had highest concentration, while under medium (L2) and high (L3) light intensities, C1 showed highest concentrations and yields. In summary, based on the results of this experiment, the best conditions to produce phycobiliproteins with better purity levels from *Nostoc* sp. were at low and medium light intensity (L1: 65 and L2: 150 μE/m$_2$s) with low or high $NaNO_3$ concentrations (C1: 0.75 and C3: 2.25 g $NaNO_3$/L).

3.1.2.2. *A. platensis*

Results of *A. platensis* cultivations show that, in average, phycocyanin was the most abundant phycobiliprotein, followed by allophycocyanin and phycoerythrin. The maximum total concentration of phycobiliproteins was obtained in L1C2 (303 mg/g DW), while the minimum in L1C3 (60 mg/g DW). For low (C1) and medium (C2) $NaHCO_3$ concentrations, lower light intensities produced more phycobiliproteins.

However, for high (C3) NaHCO$_3$ concentrations, higher light conditions produced more phycobiliproteins (Figure 3). Nevertheless, when biomass production is considered, higher light intensities led to higher phycobiliproteins yield in all cases, which was also reported in previous studies (Castro et al., 2015; Markou et al., 2012). Regarding the influence of NaHCO$_3$ in the growth medium, higher concentrations led to lower phycobiliproteins content (Sharma et al., 2014). That explains the lowest phycobiliproteins concentration and production in L1C3 (60 mg/g DW, from which 23 mg PC/g DW) and L2C3 (152 mg/g DW, from which 55 mg PC/g DW) since these trials had the less favorable condition among all others, i.e. low phycobiliproteins content due to high carbon content and low biomass production due to low light conditions. The absorbance spectra of the crude extracts also show that L1C3 and L2C3 obtained lower absorbance for phycocyanin ($\lambda = 615$ nm). An interesting finding is that these two trials also show higher absorbance than others in the range of $\lambda = 420\text{-}440$ and 660-680 nm, which suggest the presence of chlorophyll *a*. In this sense, considering that absorbance of allophycocyanin is measured at $\lambda = 652$ nm and overlaps with chlorophyll *a*, the content of allophycocyanin might have been overestimated for these trials, especially after observing that in these two cases only, the content of allophycocyanin is higher than phycocyanin. To sum up, based on the results of this experiment, the best conditions to produce phycobiliproteins with better purity levels from *A. platensis* were at medium and high light intensities (L2: 150 and L3: 230 µE/m$_2$s) combined with low and medium NaHCO$_3$ concentrations (C1: 8.4 and C2: 16.8 g/L).

3.1.2.3. *P. purpureum*

Results of *P. purpureum* show that, in average, phycoerythrin was the most abundant phycobiliprotein, followed by phycocyanin and allophycocyanin. Similarly, reducing light conditions led to higher concentrations of phycobiliproteins (Figure 3), with maximum total concentration obtained for L1C3 (93 mg/g DW) and minimum for L3C2 (29 mg/g DW). However, when biomass production and purity are considered, higher productions of phycobiliproteins were obtained in medium intensity (L2). Likewise, Sosa-Hernández et al. (2019) observed highest values of phycoerythrin content under light intensities of 65 and 100 µE $m^{-2}$ $s^{-1}$ (3.10 and 2.71 mg/g DW, respectively), which are similar to the low (L1: 65 µE/$m_2$s) and medium (L2: 150 µE/$m_2$s) light intensities in the present study. Previous studies have also reported that high light conditions contributed to *P. purpureum* biomass accumulation, but it was adverse for biosynthesis of valuable compounds, such as phycobiliproteins, arachidonic acid and total fatty acids (Guihéneuf and Stengel, 2015; Su et al., 2016). Regarding the influence of salinity levels, high NaCl concentration (C3) led to higher phycobiliproteins concentrations, especially for low (L1) and medium (L2) light intensity. Kathiresan et al. (2007) reported similar results, in which high salinity concentration (29.62 g/L) would lead to maximum phycobiliproteins production. In summary, based on the results of this experiment, the best conditions to produce phycobiliproteins with better purity levels from *P. purpureum* were at medium light intensity (L2: 150 µE/$m_2$s) with medium and high salinity concentrations (C2: 19.51 and C3: 39.02 g sea salt/L).

3.1.2.4. *Definition of conditions for the wastewater treatment experiment*

In order to select the optimal conditions for the following experiment (wastewater treatment), results were compared in terms of nutrients removal and phycobiliproteins production. For *Nostoc* sp., low (L1) and medium (L2) light conditions were optimal, but although lower light conditions (L1) reached highest phycobiliproteins yields, medium light conditions (L2) had higher purity and higher nutrients removal rates. For *A. platensis* and *P. purpurem,* medium (L2) and high (L3) light conditions were optimal. L3 led to higher TIN removal in *A. platensis* and *P. purpurem*, but L2 led to high phycobiliproteins productions with high purity while promoting similar nutrients removal compared to L3. Hence, medium light conditions (L2) was selected for the following experiment. Regarding medium composition for *Nostoc* sp., low (C1) or high (C3) concentrations of $NaNO_3$ were optimal, but removal efficiencies of C1 were much higher since C3 were limited by phosphorus. Hence, concentrations (as well as N:P ratio) similar to C1 were selected for cultivating *Nostoc* sp. in the following experiment. *A. platensis* showed higher nutrients removal in C3 and better phycobiliproteins productions in C1 and C2, so conductivities similar to C2 and C3 were selected. Finally, for *P. purpureum,* although C3 resulted in higher nutrients removal rates and phycobiliproteins, C2 and C1 were selected for the wastewater treatment experiment to avoid adding salts and avoid very low nutrients concentrations, since ASW portion would have to be much higher compared to wastewater. This way, simple and low-cost composition of growth medium using (real) wastewater mixed with seawater were evaluated, which would be more realistic in a full-scale implementation.

### *3.2. Wastewater treatment experiment*

The set-up of the wastewater treatment experiment was defined based on the results of biomass growth and phycobiliproteins production of *Nostoc* sp., *A. platensis* and *P. purpureum* studied during the optimization experiment. Five photobioreactors under medium light intensity (L2) were fed with industrial wastewater mixed with: i) standard medium BG-11 for *Nostoc* sp., at ratio 50% WW + 50% BG-11, which represented similar concentrations of C1; and ii) ASW for saline species *A. platensis* and *P. purpureum*, at two different volume ratios: 50% WW + 50% ASW and 75% WW + 25% ASW, which represented similar conductivities of C3 and C2 for *A. platensis* and C2 and C1 for *P. purpureum*, respectively.

### *3.2.1. Biomass growth and nutrients removal*

3.2.1.1. *Overall wastewater treatment efficiency*

Results showed that the three species could grow well in all photobioreactors. Cultivation period ended after 10 days, when nutrients reached low concentrations and biomass growth reached stationary phase. Biomass growth during the experiment, as well as the profiles of nitrogen species and $PO_4^{3-}$-P, are shown in Figure 4. Initial and final concentrations of the photobioreactors are summarized in Table 2. Removal efficiencies in the photobioreactors ranged from 45% (N-50%WW) to 84% (A-75%WW) for sCOD, 89% (P-75%WW) to 99% (A-50%WW) for TIN and 81% (A-75%WW) to 100% (N-50%WW and P-50%WW) for $PO_4^{3-}$-P, suggesting that the three species can potentially be applied for wastewater treatment.

Results on wastewater treatment efficiencies and specific growth rate in the present study are comparable to those previously reported on cultivation of similar microalgae in wastewaters. The observed nutrients removal efficiencies varied depending on the media composition and environmental conditions such as the influent concentrations, light conditions, N/P ratio, cultivation mode, and microalgae species. *Nostoc* sp. has been reported to efficiently treat municipal wastewater with removal efficiencies higher than 90% of NH$_4$-N and up to 60% phosphorus (El-Sheekh et al., 2014; Sharma and Khan, 2013) and to reduce COD and BOD of acidic textile effluent diluted in BG-11 medium, by 32 and 55% respectively (Talukder et al., 2015). Zhou et al. (2017) reported very similar results of nutrients removal (95% of NH$_4$-N and higher than 90% of PO$_4^{3-}$-P) by *A. platensis* grown in synthetic toilet flushing wastewater (using seawater) mixed with washing wastewater (using freshwater), in experiments with similar initial nutrients concentrations. Likewise, Chaiklahan et al. (2010) reached comparable removal of TIN by 89% and phosphorus by 57% treating swine wastewater in a semi-continuous culture mode. Regarding the use of *P. purpureum* for wastewater treatment, no other study has been found in literature.

It is important to note that TIN removal in all photobioreactors was done not only by active biomass uptake, but also by ammonia stripping (as pH ranged from 8.62 and 9.95) and denitrification. Based on the total nitrogen balance, the amount of TIN assumed to be removed by stripping or denitrification during the cultivation period was 19.6 (N-50%WW), 13.5 (A-50%WW), 26.2 (A-75%WW), 12.1 (P-50%WW) and 20.1 (P-75%WW) mg/L. As the pH of all photobioreactors was not significantly different (p-value = 0.814), thus assuming that loss through volatilization is similar in all photobioreactors, higher removal of nitrogen in N-50%WW, A-75%WW and P-75%WW might be related to higher nitrification-denitrification activity. The higher values of NO$_3$-N and NO$_2$-N in these photobioreactors, compared to A-50%WW to P-50%WW, respectively, corroborate this assumption (Table 2). TIN removal by ammonia volatilization and nitrification-denitrification was also reported in other studies cultivating *A. platensis* in wastewater (Chaiklahan et al., 2010; Zhou et al., 2017).

Overall, the results showed that the three species could grow well in all photobioreactors while efficiently removing sCOD and nutrients below discharge limits for this effluent (EEA, 2019).

*3.2.1.2. Comparison between saline species (A. platensis and P. purpureum)*

Specifically, the saline species are further compared, as they were both cultivated in the two ratios of wastewater (50% and 75% WW). First, the performance of individual species is compared in both ratios of wastewater. Later, the performances of both species are compared treating the same influent.

Biomass concentration of *A. platensis* reached 498 mg/L in A-50%WW, while in A-75%WW the concentration reached 614 mg/L, which was 23% higher (although not significantly different, p-value = 0.794). These results show that *A. platensis* could grow better in the photobioreactor with higher portion of wastewater, most probably due to the higher nutrients concentrations. The lower salinity could also have induced better growth, as previous studies suggested that higher salinity reduced *A. platensis* growth in wastewater (Duangsri and Satirapipathkul, 2011; Volkmann et al., 2008; Zhou et al., 2017).

Biomass concentration of *P. purpureum* reached 378 mg/L in P-50%WW and 496 mg/L P-75%WW at the end of the cultivation period. Similarly, the second produced 31%, although not significantly (p-value = 0.202), higher biomass concentrations at the end of the batch cultivation. These results show that *P. purpureum* could grow better in the photobioreactor with higher portion of wastewater, most probably because the influent concentrations were higher.

Although *A. platensis* produced more biomass than *P. purpureum*, there was no significant difference neither between A-50%WW and P-50%WW (p-value = 0.148) nor between A-75%WW and P-75%WW (p-value = 0.613). Likewise, in terms of treatment efficiency, despite *A. platensis* cultivations had higher removal rates than *P. purpureum* for both 50%WW and 75%WW, no significant difference was observed between TIN, $PO_4^{3-}$-P and COD removal rates (p-values ranging from 0.194 to 0.836). These results suggest that both species could be applied for efficient wastewater treatment.

### 3.2.2. *Phycobiliproteins*

The concentrations and production of phycobiliproteins extracted from the biomass grown during the wastewater treatment experiment are shown in **Error! Reference source not found.**. The concentrations represent the amount of phycobiliproteins per dry weight, while the production represents the amount of phycobiliprotein per unit of volume and time in each trial, i.e. considering the content and biomass production during this period. Further discussion on the phycobiliproteins content and production from biomass grown in wastewater are provided in this section.

Among all photobioreactors, the highest total phycobiliproteins was observed in N-50%WW with 179 mg PBP/g DW, while the lowest in P-50%WW with 36 mg PBP/g DW. However, when biomass production is considered, photobioreactors of *A. platensis* (A-50%WW and A-75%WW) had highest production of phycobiliproteins. It is important to note that the absorbance spectra of the crude extracts of *Nostoc* sp. and *A. platensis* showed absorbance in the range of λ = 420-440 and 660-680 nm, which suggest the presence of chlorophyll *a*. In this sense, as previously described in Section 3.1.2, the content of allophycocyanin might have been overestimated for N-50%WW, A-50%WW and A-75%WW. Overall, the results suggest that the three species could produce phycobiliproteins while treating industrial wastewater.

3.2.2.1. *Comparison between saline species (A. platensis and P. purpureum)*
Total phycobiliproteins concentrations in A-50%WW were 13%, although not significantly (p-value = 0.702), higher than in A-75%WW. However, when biomass production is considered, A-75%WW showed 9% higher production than A-50%WW (p-value=0.689), but with lower purity factors. Regarding *P. purpureum*, total phycobiliproteins concentrations in P-75%WW were 31% significantly higher than in P-50%WW (1.2 x 10$^{-4}$). When biomass production is considered the discrepancy is even higher, with P-75%WW showing 72% higher phycobiliproteins production than P-50%WW (p-value=2.5 x 10$^{-6}$), but with lower phycoerythrin purity factors. The results suggest that A-50%WW and P-75%WW would be the appropriate conditions for *A. platensis* and *P. purpureum* to maximize phycobiliproteins production.
Nitrogen is an important source for both the biomass and phycobiliproteins production. Zhao et al. (2017) demonstrated that a nitrate starvation led to the decline of photosynthetic performance, which is directly related to photosynthetic pigments such as chlorophyll and phycobiliproteins. Considering that the nutrients were depleted in all photobioreactors by the end of the experimental period, not only biomass but also phycobiliproteins production could possibly be improved by using higher influent nutrient concentrations or (semi) continuous culture mode.

*3.2.2.2. Comparison of phycobiliproteins production with synthetic medium and real wastewater*

The concentration of total phycobiliproteins observed in N-50%WW (179 mg PBP/g DW) was higher than in the cultivation of synthetic medium under similar conditions L2C1 (127 mg PBP/g DW). However, purity of phycocyanin, allophycocyanin and phycoerythrin were lower in N-50%WW (0.56, 0.32 and 0.35, respectively) than in L2C1 (0.91, 0.43 and 0.47). In addition, as previously mentioned, there might be an overestimation on this value after analyzing the absorbance spectra of their crude extracts. Considering that TIN removal is associated with the biosynthesis of phycobiliproteins, the production from *Nostoc* sp. grown in wastewater was much lower than in standard growth medium. The productivity of total phycobiliproteins in N-50%WW was 0.44 mg PBP/mg TIN removed by biomass uptake (i.e. disregarding the TIN removed stripping or denitrification) (average production of 5.8 mg PBP/Ld), while in L2C1 the productivity was 1.8 mg PBP/mg TIN. Although initial nutrients concentrations in N-50%WW were 27% and 15% higher than in L2C1 (128 mg TIN/L and 8 mg $PO_4^{3-}$-P/L), respectively, total phycobiliproteins production in L2C1 was 64% higher. These results showed that the productivity of phycobiliproteins from *Nostoc* sp. biomass grown in wastewater were somewhat lower than in standard growth medium, but it still shows promise as an alternative to recover resources from wastewater. In addition, further optimization could be explored in terms of type of wastewater and operational conditions in order to enhance the productivity.

The production from *A. platensis* biomass grown in wastewater was comparable to biomass grown in standard growth medium. The productivity of total phycobiliproteins in A-50%WW was 2.0 mg PBP/mg TIN (average production of 6.4 mg PBP/Ld).

Similar conditions were imposed during optimization experiment L2C3, which resulted in similar productivity of 1.3 mg PBP/mg TIN. The phycobiliproteins production in this case was 2.4-fold higher, but it is important to mention that initial nutrients concentrations (259 mg TIN/L and 88 mg $PO_4^{3-}$-P/L) were 5.5-fold and 19.4-fold higher than in A-50%WW, respectively. Likewise, the productivity of total phycobiliproteins in A-75%WW was 1.8 mg PBP/mg TIN (production of 7.0 mg PBP/Ld). During the optimization experiment L2C2, similar conditions were imposed and comparable productivity of 1.9 mg PBP/mg TIN was observed. In this case, phycobiliproteins production was 2.7-fold higher (19 mg PBP/Ld), also due to higher initial nutrients concentrations, which were 3.5-fold and 9.3-fold higher than in A-75%WW, respectively.

The production from *P. purpureum* biomass grown in wastewater was comparable to biomass grown in standard growth medium. The productivity of total phycobiliproteins in P-50%WW was 0.41 mg PBP/mg TIN (average production of 1.3 mg PBP/Ld). Similar conditions were imposed during optimization experiment L2C2, which resulted in productivity of 0.75 mg PBP/mg TIN. The phycobiliproteins production in L2C2 was 4.2-fold higher (5.6 mg PBP/Ld), due to higher initial nutrients concentration, which were 3.3-fold and 2.7-fold higher than in P-50%WW, respectively. Similarly, the productivity of total phycobiliproteins in P-75%WW was 0.6 mg PBP/mg TIN (production of 2.3 mg/Ld). During the optimization experiment L2C1, similar conditions were imposed, but productivity was 0.7 mg PBP/mg TIN. In this case, phycobiliproteins production was 1.6-fold higher (3.6 mg/Ld), also due to higher initial nutrients concentrations, which were 2.4-fold and 1.1-fold higher than in P-75%WW, respectively.

It is important to notice that several researchers have investigated the three species studied in this work, but almost all of them describe cultivations using standard growth medium. A few studies for *A. platensis* and no studies for *Nostoc* sp. and *P. purpureum* addressed optimization of phycobiliproteins production from biomass grown in wastewater. Results on phycobiliproteins produced in the present study are comparable to those previously reported in studies using microalgae grown in wastewater. Hultberg et al. (2017) cultivated *A. platensis* in anaerobic digestate effluent diluted (6%) in carbonate buffer and obtained higher phycocyanin (86.2 mg/g DW) and allophycocyanin (41.3 mg/g DW) compared to biomass grown in synthetic medium. Khatoon et al. (2018) observed higher phycobiliproteins production (175 mg PBP/g DW) than this study (114 – 129 mg PBP/g DW), in spite of much lower nutrients concentrations, suggesting that *P. mucicola* might produce more phycobiliproteins than *A. platensis*. Wood et al. (2015) demonstrated that wastewater from oil and natural gas extraction, amended with 3 g $NaNO_3$/L and 0.5 g $K_2HPO_4$/L, could support growth of a cyanobacterial consortium, mainly composed by Oscillatoriales, but produced phycocyanin yields (16.9 mg/g DW) much lower than in this study. This might be explained by the mixed culture used (instead of unialgal cultures applied in this work) or the distinct configuration used (rotating algal biofilm reactor). Similarly, Van Den Hende et al. (2016b) carried out a study treating the UASB effluent from the same source of this work and reported lower phycobiliproteins contents and purity levels using one-step extraction, which might also be due to the mixed culture applied (containing cyanobacteria *Geminocystis* sp. and diatoms).

Although comparisons with the phycobiliproteins productions in standard growth media and wastewater have been discussed in this section, it is noteworthy to highlight that these comparisons considered only the parameters evaluated in both experiments, i.e. how light conditions and growth medium concentrations, combined with available nutrients could affect the content and production of phycobiliproteins. Hence, in the future, further optimization can be performed in terms of other medium components and operation conditions, in order to have a holistic understanding on the performance of these three species in wastewater, especially when aiming for biomass valorization (phycobiliproteins production and purity factor).

The results from the present study described the potential of applying cyanobacteria, such as *Nostoc* sp. and *A. platensis,* and red microalgae, such as *P. purpureum*, for combining wastewater treatment and resources recovery. This work has focused on the phycobiliproteins production from these two species, but several researchers highlights the potential of microalgae as a factory for high-value compounds, since its composition can be modified according to operational variables to obtain protein enrichment, such as reducing residence times (Rebolloso Fuentes et al., 2000), semi continuous or continuous cultivation mode (Guihéneuf and Stengel, 2015) or mixotrophic growth (Fábregas et al., 1999).

Finally, this study showed that industrial wastewater could be applied as a medium in order to not only promote biomass growth and cleaner water, but also to reduce typical high costs to produce valuable compounds from microalgae, such as pigments. The wastewater was an effluent from a food processing company, so no potential contaminants were detected. This was already expected, since the company has to comply with high standards, according to food safety regulations. However, it is important to mention that for further development of the process proposed in this study, biosafety concerns have to be considered. Depending on the desired application of the pigments, further analyses should be done to assure that the bioproducts do not present any potential risk. In this particular study, the pigments extracted will be used for a project of art and design, highlighting the applicability of such bioproducts as natural textile dyes (Ferrándiz et al., 2016; Moldovan et al., 2017) and raising no major concerns regarding potential risks by contaminants. Future research is thus encouraged in order to address the current challenges, such as cultivation systems and extraction methods. This way, once technical feasibility and economic viability of this concept is ensured, its further development as a resource recovery solution should move towards regulations analysis and decision-making processes.

**4. Conclusions**

This study suggested that, in general, light conditions had more influence on biomass growth and phycobiliproteins production than the medium composition in the cultivations of *Nostoc* sp., *A. platensis* and *P. purpureum*. The three species showed efficient treatment of industrial wastewater reaching high COD and nutrients removal, while successfully biosynthesizing high-value compounds in their biomass. This study encourages further investigations on the feasibility of this process, as well as research developments with a holistic approach to explore other synergetic opportunities associated with the nexus of water and sustainable resource recovery processes.

*E-supplementary data of this work can be found in online version of the paper.*

**Acknowledgements**

This research was funded by the European Union's Horizon 2020 research and innovation program under the Marie Skłodowska-Curie grant agreement No 676070 (SuPER-W). This communication reflects only the author's view and the Research Executive Agency of the EU is not responsible for any use that may be made of the information it contains. Authors are grateful to the Spanish Ministry of Science, Innovation and Universities (MCIU), Research National Agency (AEI), and European Regional Development Fund (FEDER) for the AL4BIO project (RTI2018-099495-B-C21). Marianna Garfí is grateful to the Spanish Ministry of Economy and Competitiveness (RYC-2016-20059). The authors acknowledge Juan Luis Gómez Pinchetti from the Spanish Bank of Algae for his contribution providing the microalgae species *Nostoc* sp.; Dave Manhaeghe, Pieter-Jan De Buyck and Pieter Knockaert from Ghent University (campus Kortrijk) for their help with the experimental set-up; Marijn Alliet from Alpro for kindly providing the wastewater samples; and Javier Carretero from UPC for the assistance with the recovery of pigments.

**Table 1.** Varying growth conditions of light intensity and medium concentration (with respective conductivity) tested for *Nostoc* sp., *A. platensis* and *P. purpureum* cultivation.

| Trial | Light intensity ($\mu E/m^2 s$) | *Nostoc* sp. | | *A. platensis* | | *P. purpureum* | |
|---|---|---|---|---|---|---|---|
| | | $NaNO_3$ (g/L) | Conductivity (mS/cm) | $NaHCO_3$ (g/L) | Conductivity (mS/cm) | Sea salt (g/L) | Conductivity (mS/cm) |
| L1C1 | 65 ± 6 | 0.75 | 1.18 ± 0.01 | 8.4 | 13.07 ± 0.01 | 9.76 | 16.08 ± 0.02 |
| L1C2 | 65 ± 6 | 1.5 | 2.22 ± 0.02 | 16.8 | 19.29 ± 0.01 | 19.51 | 29.01 ± 0.01 |
| L1C3 | 65 ± 6 | 2.25 | 3.16 ± 0.02 | 25.2 | 25.13 ± 0.03 | 39.02 | 41.63 ± 0.09 |
| L2C1 | 150 ± 7 | 0.75 | 1.18 ± 0.01 | 8.4 | 13.07 ± 0.01 | 9.76 | 16.08 ± 0.02 |
| L2C2 | 150 ± 7 | 1.5 | 2.22 ± 0.02 | 16.8 | 19.29 ± 0.01 | 19.51 | 29.01 ± 0.01 |
| L2C3 | 150 ± 7 | 2.25 | 3.16 ± 0.02 | 25.2 | 25.13 ± 0.03 | 39.02 | 41.63 ± 0.09 |
| L3C1 | 230 ± 9 | 0.75 | 1.18 ± 0.01 | 8.4 | 13.07 ± 0.01 | 9.76 | 16.08 ± 0.02 |
| L3C2 | 230 ± 9 | 1.5 | 2.22 ± 0.02 | 16.8 | 19.29 ± 0.01 | 19.51 | 29.01 ± 0.01 |
| L3C3 | 230 ± 9 | 2.25 | 3.16 ± 0.02 | 25.2 | 25.13 ± 0.03 | 39.02 | 41.63 ± 0.09 |

**Table 2.** Average initial and final concentrations of water quality parameters measured in the cultivations of *Nostoc* sp. with 50% wastewater (N-50%WW), *A. platensis* with 50% wastewater (A-50%WW) and 75% wastewater (A-75%WW), and cultivations of *P. purpureum* with 50% wastewater (P-50%WW) and 75% wastewater (P-75%WW).

| Microalgae | | *Nostoc* sp. | | *A. platensis* | | | | *P. purpureum* | | | |
|---|---|---|---|---|---|---|---|---|---|---|---|
| Photobioreactor | | N-50%WW | | A-50%WW | | A-75%WW | | P-50%WW | | P-75%WW | |
| | | Initial | Final | Initial | Final | Initial | Final | Initial | Final | Initial | Final |
| pH | - | 7.61 ± 0.01$_a$ | 9.38 ± 0.03$_b$ | 8.94 ± 0.01$_a$ | 9.90 ± 0.01$_b$ | 8.62 ± 0.01$_a$ | 9.95 ± 0.01$_b$ | 8.94 ± 0.01$_a$ | 9.25 ± 0.01$_b$ | 8.62 ± 0.01$_a$ | 9.43 ± 0.01$_b$ |
| EC | mS/cm | 2.92 ± 0.01$_a$ | 2.82 ± 0.01$_b$ | 24.5 ± 0.06$_a$ | 23.60 ± 0.06$_b$ | 14.47 ± 0.01$_a$ | 14.97 ± 0.01$_b$ | 24.50 ± 0.06$_a$ | 23.90 ± 0.06$_b$ | 14.47 ± 0.01$_a$ | 14.20 ± 0.02$_b$ |
| sCOD | mg/L | 107 ± 2$_a$ | 59 ± 10$_b$ | 118 ± 9$_a$ | 21 ± 2$_b$ | 159 ± 3$_a$ | 26 ± 3$_b$ | 155 ± 4$_a$ | 38 ± 4$_b$ | 198 ± 12$_a$ | 41 ± 2$_b$ |
| TN* | mg/L | 205 ± 2$_a$ | 185.4 ± 3.4$_b$ | 82.5 ± 0.3 | 69 ± 8 | 98.1 ± 1.9$_a$ | 72 ± 1$_b$ | 75.5 ± 0.3 | 63 ± 14 | 112.5 ± 0.3 | 92 ± 8 |
| $NH_4^+$-N | mg/L | 44.4 ± 1.7$_a$ | 0.05 ± 0.01$_b$ | 38.4 ± 0.2$_a$ | 0.0 ± 0.0$_b$ | 61.5 ± 0.3$_a$ | 0.0 ± 0.0$_b$ | 43.6 ± 0.1$_a$ | 0.0 ± 0.0$_b$ | 58.7 ± 0.4$_a$ | 4.24 ± 0.01$_b$ |
| $NO_3^-$-N | mg/L | 118 ± 2$_a$ | 9.9 ± 0.8$_b$ | 8.35 ± 0.04$_a$ | 0.60 ± 0.8$_b$ | 11.9 ± 0.1$_a$ | 7.52 ± 0.07$_b$ | 3.85 ± 0.02$_a$ | 2.2 ± 0.1$_b$ | 8.25 ± 0.02$_a$ | 3.4 ± 0.1$_b$ |
| $NO_2^-$-N | mg/L | 0.10 ± 0.01$_a$ | 2.52 ± 0.05$_b$ | 0.09 ± 0.04$_a$ | 0.61 ± 0.01$_b$ | 0.14 ± 0.03$_a$ | 2.43 ± 0.04$_b$ | 0.09 ± 0.02$_a$ | 0.45 ± 0.02$_b$ | 0.14 ± 0.01$_a$ | 0.51 ± 0.01$_b$ |
| TP* | mg/L | 205 ± 2 | 14.0 ± 0.1 | 18.5 ± 0.1 | 17.4 ± 0.6 | 22.5 ± 0.2 | 19 ± 1 | 12.6 ± 0.7 | 11.1 ± 0.7 | 14.9 ± 0.2 | 12.6 ± 1.1 |
| $PO_4^{3-}$-P | mg/L | 9.11 ± 0.04$_a$ | 0.04 ± 0.01$_b$ | 4.5 ± 0.7$_a$ | 0.4 ± 0.0$_b$ | 9.7 ± 0.1$_a$ | 1.8 ± 0.2$_b$ | 5.24 ± 0.03$_a$ | 0.02 ± 0.01$_b$ | 11.6 ± 0.2$_a$ | 0.2 ± 0.1$_b$ |

Acronyms: EC (electrical conductivity); sCOD (soluble chemical oxygen demand); TN (total nitrogen); TP (total phosphorus).
*TN and TP were measured in the mixed liquor (influent with inoculum).
$_{a,b}$: Letters indicate a significant difference (α=0.05) between influent and effluent concentrations after Fisher test.

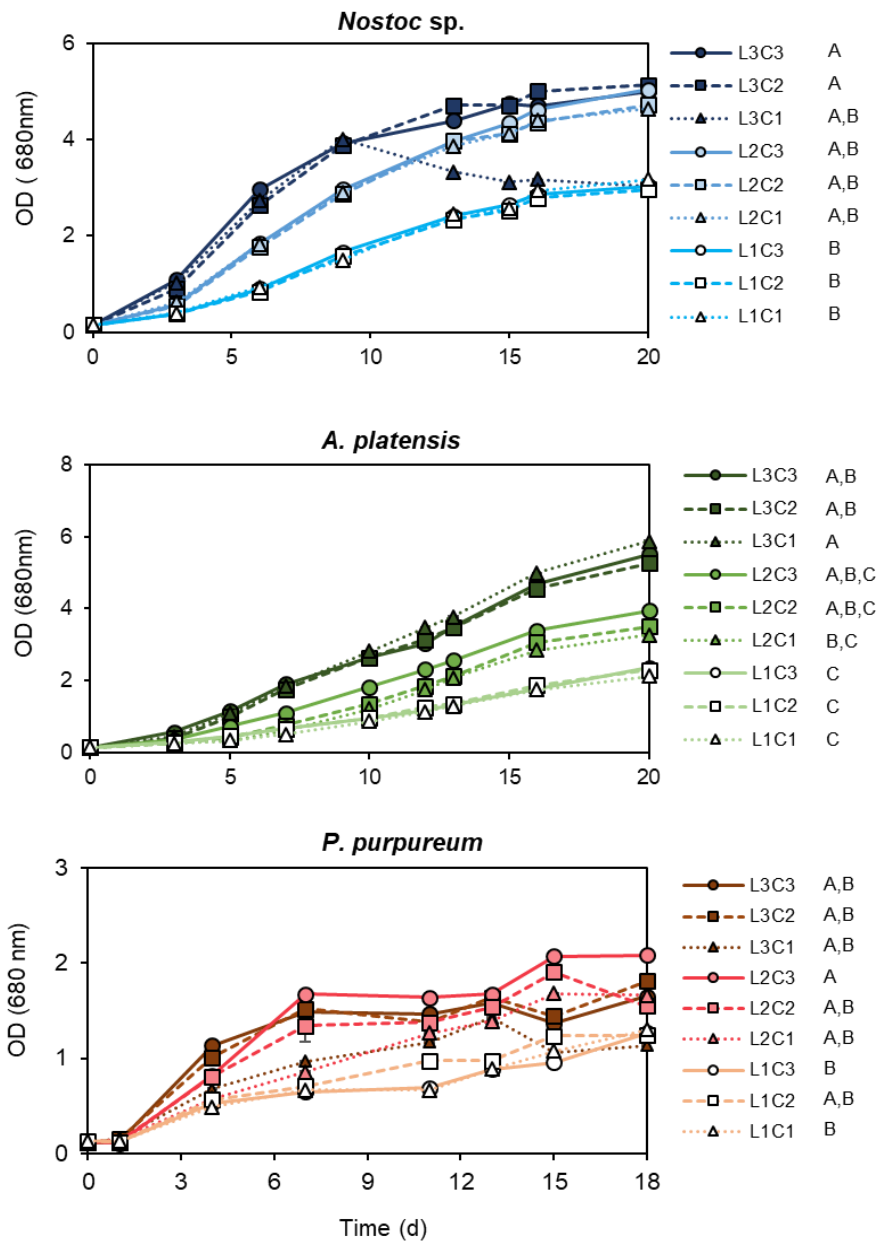

**Figure 1**. Biomass growth of *Nostoc* sp., *A. platensis* and *P. purpureum* during the optimization experiment, measured as optical density (OD) at 680nm. Biomass growth was monitored under different conditions of low (C1), medium (C2) and high (C3) concentration of the main ingredient in standard growth medium ($NaNO_3$ for *Nostoc* sp., $NaHCO_3$ for *A. platensis* and NaCl as sea salt for *P. purpureum*) and low (L1), medium (L2) and high (L3) light intensity. Letters A, B and C indicate a significant difference ($\alpha=0.05$) among trials after Fisher test.

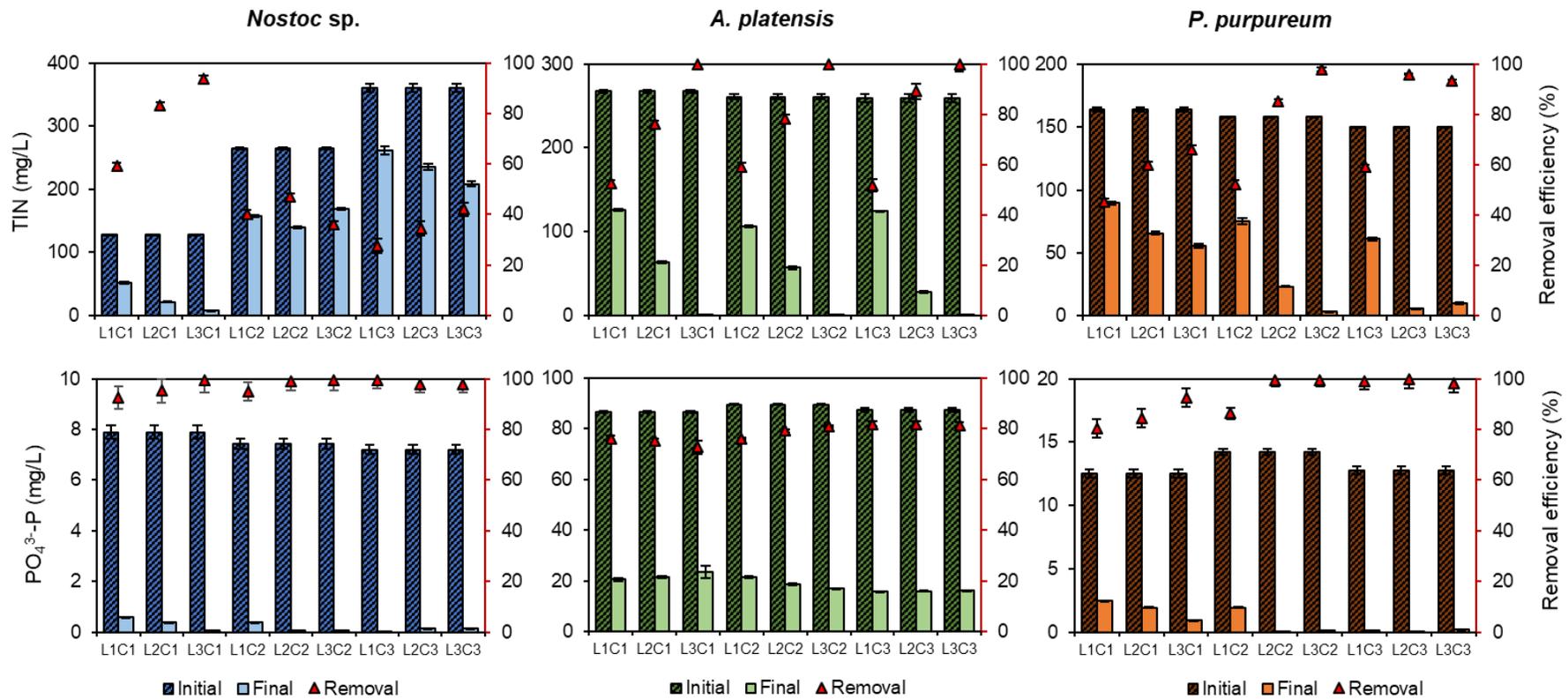

**Figure 2**. Initial and final concentrations of total inorganic nitrogen (TIN) and $PO_4^{3-}$-P, and respective removal efficiency (%), measured in the growth media of *Nostoc* sp., *A. platensis* and *P. purpureum* during the optimization experiment. Variations of these parameters were monitored under different conditions of low (C1), medium (C2) and high (C3) concentration of the main ingredient in standard growth medium ($NaNO_3$ for *Nostoc* sp., $NaHCO_3$ for *A. platensis* and NaCl as sea salt for *P. purpureum*) and low (L1), medium (L2) and high (L3) light intensity.

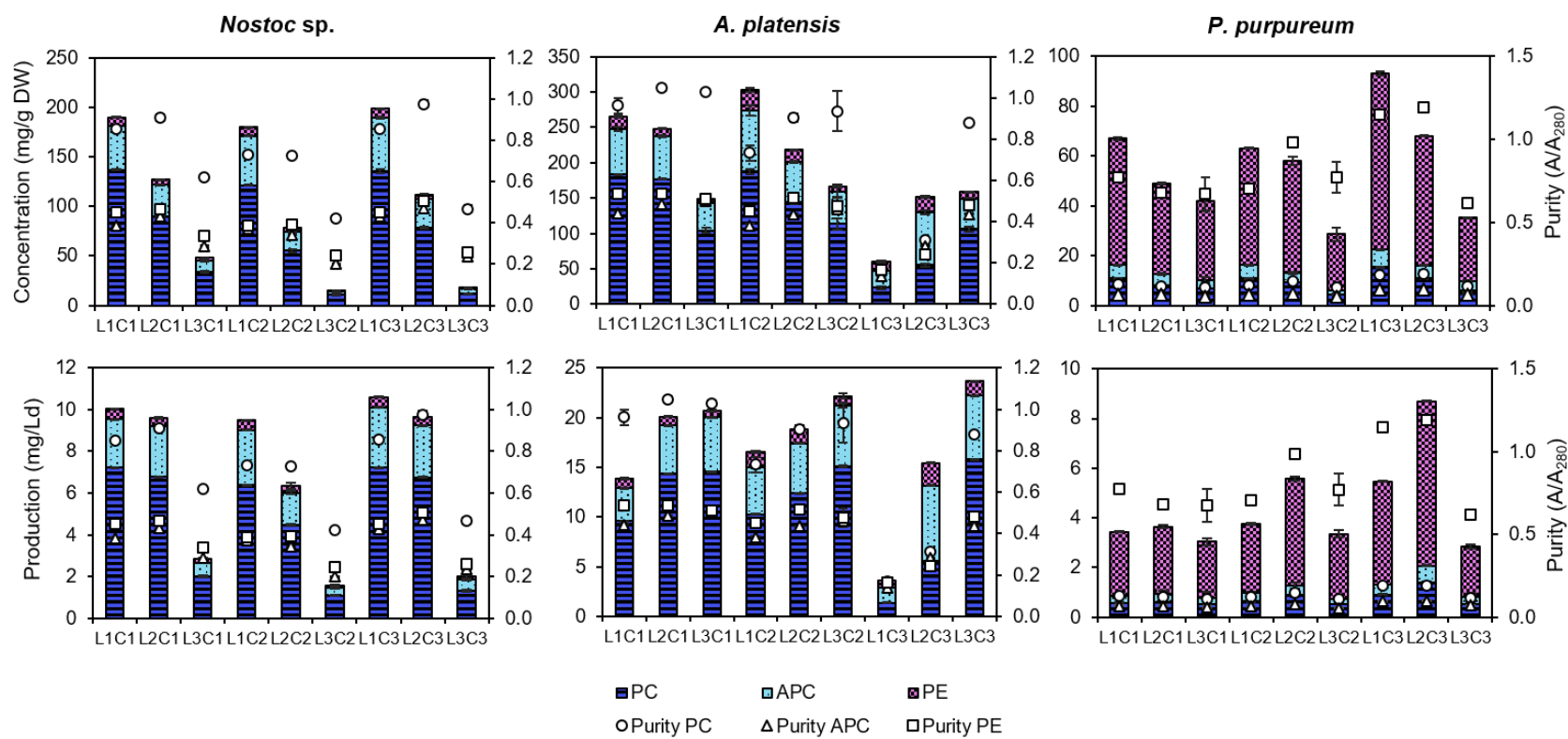

**Figure 3**. Average concentration and production of phycocyanin (PC), allophycocyanin (APC) and phycoerythrin (PE), with respective purity grades, extracted from *Nostoc* sp., *A. platensis* and *P. purpureum* biomass grown during the optimization experiment. Biomass growth was monitored under different conditions of low (C1), medium (C2) and high (C3) concentration of the main ingredient in standard growth medium ($NaNO_3$ for *Nostoc* sp., $NaHCO_3$ for *A. platensis* and NaCl as sea salt for *P. purpureum*) and low (L1), medium (L2) and high (L3) light intensity.

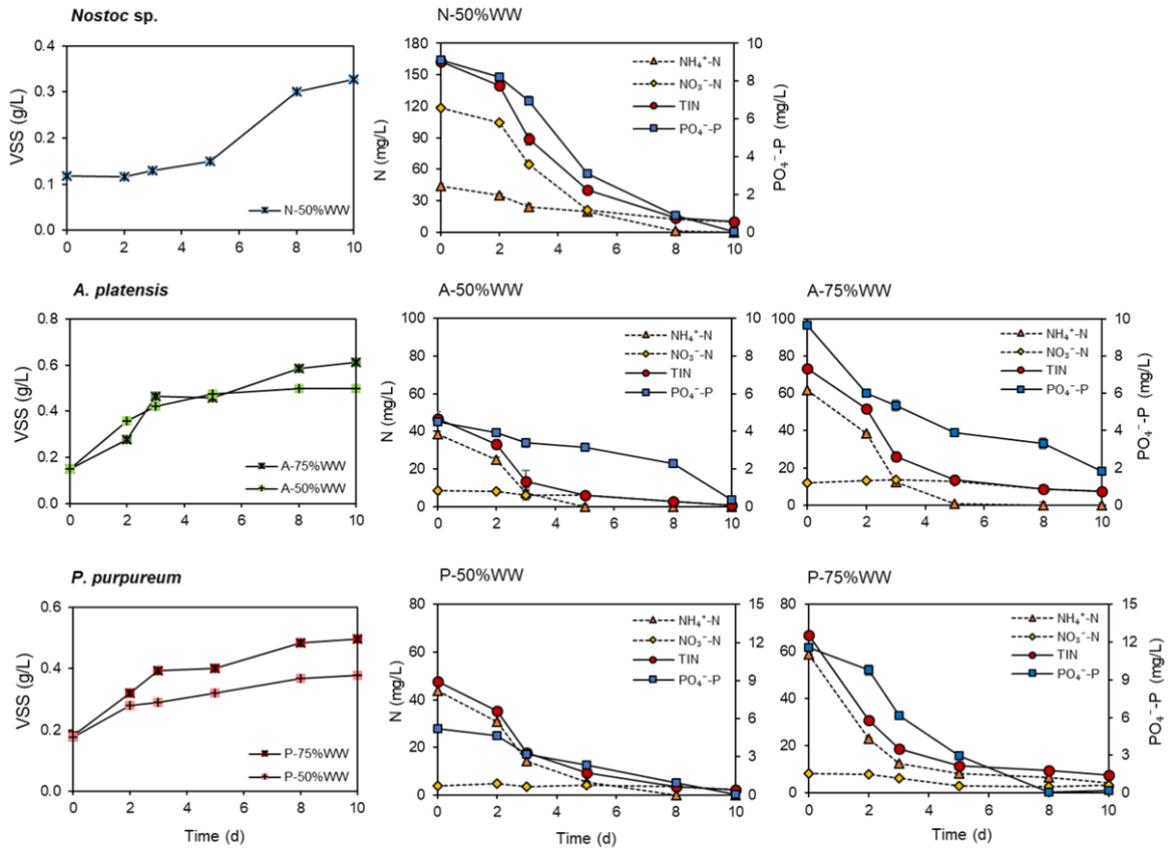

**Figure 4**. Biomass growth, measured as concentration of volatile suspended solids (VSS), and concentrations of nitrogen (N), as ammonium ($NH_4^+$-N), nitrate ($NO_3^-$-N) and total inorganic nitrogen (TIN), and $PO_4^{3-}$-P during the cultivation in different ratios of wastewater (WW) of freshwater species *Nostoc* sp. with 50% wastewater (N-50% WW), as well as saline species *A. platensis* with 50% wastewater (A-50%WW) and 75% wastewater (A-75%WW), and *P. purpureum* with 50% wastewater (P-50%WW) and 75% wastewater (P-75%WW).

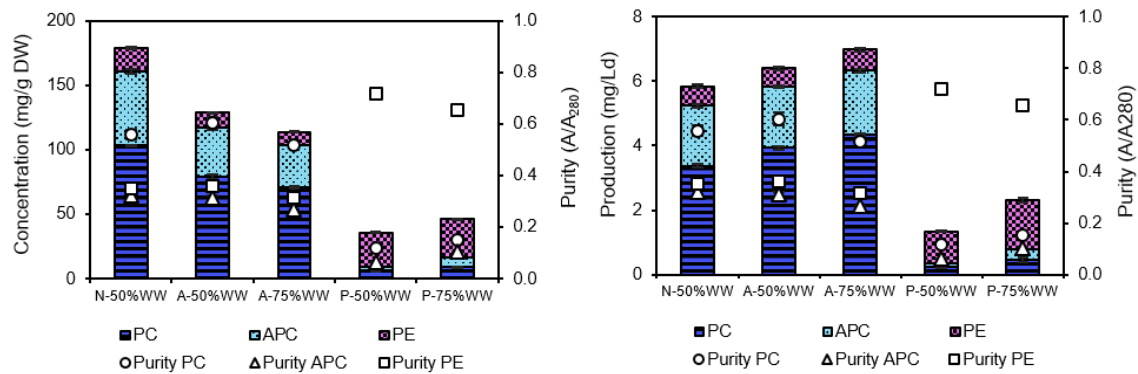

**Figure 5**. Average concentration and production of phycocyanin (PC), allophycocyanin (APC) and phycoerythrin (PE), with respective purity grades, extracted from biomass grown during the cultivation in wastewater (WW) of freshwater species *Nostoc* sp. with 50% wastewater (N-50% WW), as well as saline species *A. platensis* with 50% wastewater (A-50%WW) and 75% wastewater (A-75%WW), and *P. purpureum* with 50% wastewater (P-50%WW) and 75% wastewater (P-75%WW).